# Giant Bandgap Pulsation Driven by Hotspot Breathing Phonons in a Flat-Band Solid

Wenjie Liu,[1] Huaxin Wu,[1] and Jiyang Fan[1,2*]

[1]School of Physics, Southeast University, Nanjing 211189, P. R. China
[2]Key Laboratory of Quantum Materials and Devices of Ministry of Education, Southeast University, Nanjing 211189, P. R. China
*Contact author: jyfan@seu.edu.cn

**Abstract:** The electronic bandgap of solids is conventionally viewed as a static property at a given temperature, with only weak and stochastic thermal fluctuations under equilibrium conditions. Here, using *ab initio* molecular dynamics and first-principles electron–phonon calculations, we reveal a pronounced room-temperature bandgap pulsation at ~7.8 THz in a perovskite-like flat-band solid, with a maximum peak-to-peak variation approaching 0.95 eV. This behavior originates from a dual selection mechanism: the $A_{1g}$-like breathing branch couples much more strongly to the flat conduction-band edge than other phonon branches, while real-space phase selectivity distinguishes its hotspot Γ-point and finite-q components. Although finite-q modes retain appreciable microscopic coupling, their intercell phase shifts produce smaller-amplitude shorter-recurrence-period responses, leaving the unit-cell-synchronous Γ-point $A_{1g}$ component to dominate the fundamental-period bandgap pulsation. The resulting band-edge dynamics further modulates the optical response on femtosecond timescales. These findings demonstrate that an unexpectedly ordered electronic response can emerge from intrinsically disordered thermal lattice fluctuations.

The electronic bandgap is one of the most fundamental quantities in condensed-matter physics, governing the electrical and optical properties of materials [1]. At a given temperature, the bandgap of a solid is generally regarded as a static quantity, exhibiting only small and irregular thermal fluctuations under equilibrium conditions, although it may vary substantially with temperature [2–7]. This conventional picture originates from the nature of lattice vibrations at finite temperature: atomic motion arises from the simultaneous excitation of many phonon modes with different frequencies and phases, leading to intrinsically disordered lattice fluctuations. As a result, phonon-induced bandgap variations are expected to be noise-like and to average out in time, reinforcing the static-bandgap picture that has long underpinned the design and interpretation of optoelectronic devices. However, this perspective may be fundamentally incomplete. Lattice vibrations are inherently dynamical on the femtosecond timescale, raising the possibility that a purely static view may overlook important emergent phenomena. This issue becomes particularly relevant in materials with strong electron–phonon coupling and narrow electronic bandwidths, such as flat-band systems [8–11] and organic–inorganic hybrid semiconductors [12–14], where localized electronic states and soft lattices can amplify the impact of specific lattice distortions. While prior studies have revealed rich phonon effects on transport [8–10,15], excitons [16–20], and polarons [21,22], often in excited or laser-driven states, a pivotal question remains: can intrinsic thermal vibrations alone, without any external drive, generate nontrivial time-dependent electronic behavior under equilibrium conditions?

Here, we report the discovery of giant terahertz-frequency bandgap dynamics in a model flat-band solid at room temperature. Using ab initio molecular dynamics (AIMD) [2–5], we observe spontaneous periodic modulation of the bandgap and concomitant ultrafast evolution of optical absorption in an inorganic–organic hybrid perovskite-like crystal [23]. We trace this behavior to a dual selection mechanism. First, the $A_{1g}$-like breathing branch couples much more strongly to the flat conduction-band edge than other phonon branches. Second, within this branch, real-space phase selectivity distinguishes the unit-cell-synchronous Γ-point hotspot component from finite-q modes: intercell phase shifts reduce the modulation amplitude and shorten the recurrence period of finite-wavelength responses, leaving the Γ-point component to dominate the fundamental-period bandgap pulsation. Our results demonstrate that an unexpectedly ordered electronic response can emerge from intrinsically disordered thermal lattice fluctuations, without the external fields commonly required to generate or reveal quantum oscillatory phenomena [24–29]. These results further suggest that this dual selectivity, combining reciprocal-space branch selection with real-space phase selection, may provide a route toward intrinsic ultrafast electronic functionality in flat-band materials.

The DMSO crystal is a two-dimensional layered organic–inorganic hybrid semiconductor composed of hierarchically self-assembled dumbbell-like units, each consisting of two organic crown ether rings sandwiching a $[TeCl_6]^{2-}$ octahedron (via a $Cs^+$ bridge on each side), with small DMSO molecules occupying the interstitial voids (Fig. S1) [23]. Figure 1(a) shows the temporal evolution of the electronic energy levels near the band edges extracted from a 300 K AIMD trajectory using a 2×2×1 supercell containing 452 atoms, as obtained from the density functional theory (DFT) calculation (SI Appendix) [30–34]. Remarkably, the conduction band minimum (CBM) and nearby unoccupied states exhibit pronounced large-amplitude periodic oscillations, with an average period of approximately 129 fs, corresponding to a frequency of 7.8 THz. As a consequence, the fundamental bandgap undergoes substantial dynamical modulation [Fig. 1(a)], with a maximum peak-to-peak variation approaching 0.95 eV.

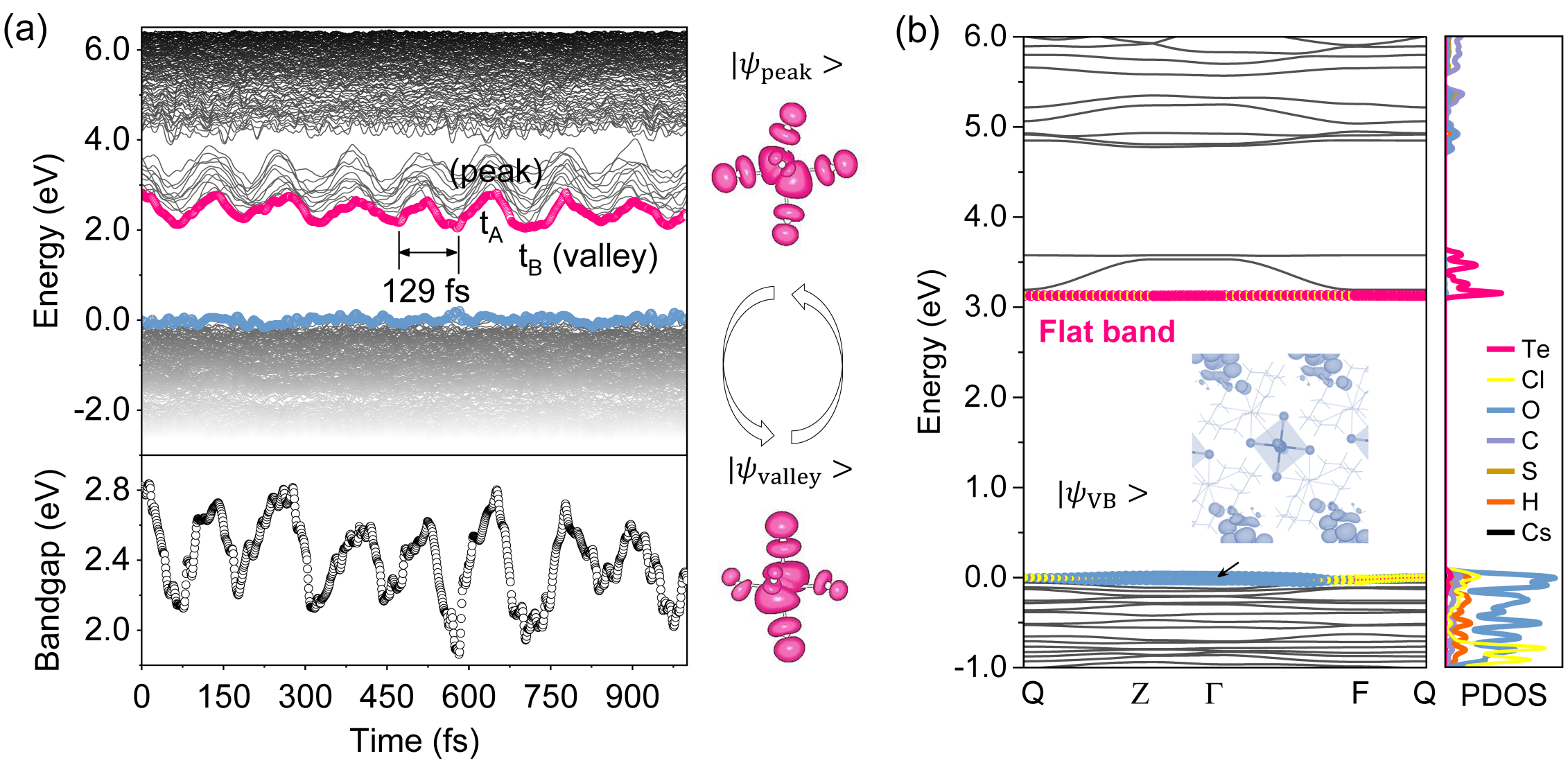

**FIG. 1. Spontaneous temporal dynamics of electronic states and band structure of the DMSO crystal.** (a) Temporal evolution of the electronic energy levels extracted from AIMD trajectories using the 2×2×1 supercell (452 atoms) at thermal equilibrium (300 K), with the average valence-band maximum (VBM) set to 0 eV. The conduction-band minimum (CBM) and the subsequent bandgap exhibit large-amplitude periodic oscillations (period: ~129 fs). The right panels show the isosurfaces of the charge density of the CBM state at two representative time points: a peak ($t_A$) and a valley ($t_B$) during the oscillation. (b) Electronic band structure and projected density of states (PDOS). Pink, yellow, and blue spheres indicate the Te, Cl, and O atomic-orbital contributions, respectively. The lowest unoccupied band features a highly localized nearly dispersionless flat band (highlighted by the pink markers).

This behavior contrasts with the weak irregular thermal fluctuations typically observed in conventional semiconductors. Comparative AIMD simulations of Si at 300 K show only much smaller irregular thermal fluctuations (Fig. S2), supporting that the pronounced pulsation in DMSO is material-specific rather than a generic feature of the AIMD procedure. Similar CBM pulsations were also observed in a N,N-dimethylformamide (DMF) (substituting for DMSO)-based hybrid material (Fig. S3), suggesting that this phenomenon may extend to related flat-band systems. To elucidate the microscopic origin of this intriguing band-edge dynamics, we first examine the static electronic structure of the DMSO crystal. The lowest unoccupied band is extremely flat across the Brillouin zone [Fig. 1(b)]. Hybrid Heyd-Scuseria-Ernzerhof (HSE06) functional calculations further confirm that this flat conduction-band character is robust beyond the PBE description (Fig. S4). The projected density of states (PDOS) shows that the CBM is dominated by Te and Cl atomic orbitals, indicating that the low-energy conduction states are primarily confined within and around the inorganic $[TeCl_6]^{2-}$ octahedra. Interestingly, the highest occupied band is also relatively flat; however, in contrast to the CBM, it is predominantly contributed by the organic molecular orbitals [Fig. 1(b)], spatially decoupled from the inorganic octahedral framework. Despite exhibiting noticeable thermal fluctuations during the AIMD evolution, the VBM does not display dynamics comparable in amplitude to that of the CBM. We further examine the real-space charge density of the CBM at the peak ($t_A$) and valley ($t_B$) [Fig. 1(a)]. At both moments, the CBM charge density remains predominantly localized on the inorganic octahedra, with its spatial extent modulated in synchrony with the energy level, motivating a detailed investigation of the lattice vibrational modes.

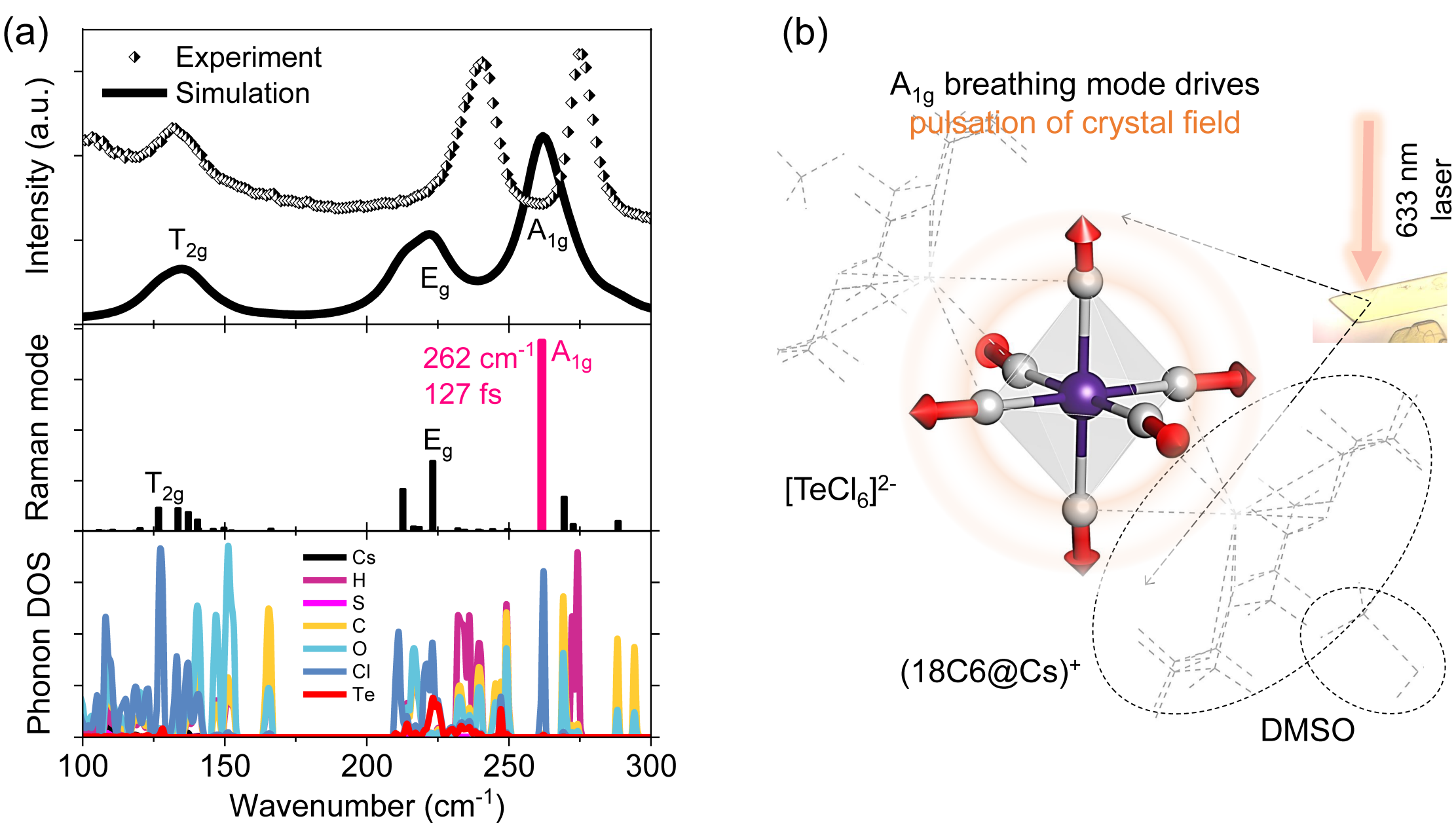

**FIG. 2. Raman spectra and $A_{1g}$-mode breathing vibration of the DMSO crystal.** (a) Measured and simulated Raman spectra of the DMSO crystal (upper panel), alongside the corresponding phonon density of states decomposed into atomic contributions (lower panel). The prominent $A_{1g}$ peak at 262 $cm^{-1}$ (corresponding to a period of 127 fs) is clearly identified. (b) Schematic illustration of the Γ-point $A_{1g}$ breathing vibration of the inorganic $[TeCl_6]^{2-}$ octahedral unit. This mode drives an in-phase expansion and contraction of the octahedral cage, periodically pulsating the local crystal field.

The lattice vibrational modes underlying this octahedron-associated electronic dynamics were investigated using Raman spectroscopy and first-principles phonon calculations. The measured Raman spectrum [Fig. 2(a)] displays three prominent peaks at 132.7, 240.5, and 274.4 $cm^{-1}$, which correspond well to the calculated Raman-active modes at 135.0 $cm^{-1}$ ($T_{2g}$), 222.0 $cm^{-1}$ ($E_g$), and 262.0 $cm^{-1}$ ($A_{1g}$), respectively. Analysis of the phonon density of states [Fig. 2(a)] together with visualization of the atomic-displacement patterns [Fig. 2(b)] [35] indicates that all three Raman-active modes are dominated by vibrations of the inorganic $[TeCl_6]^{2-}$ octahedra. Notably, although the AIMD atomic trajectories are globally irregular and nonperiodic (Fig. S5), as expected for thermal motion, the calculated $A_{1g}$ breathing mode has a period of approximately 127 fs (7.9 THz), closely matching the ~129 fs periodicity of the CBM dynamics [Fig. 1(a)]. This correspondence points to the $A_{1g}$ breathing mode as a likely microscopic origin of the observed electronic response.

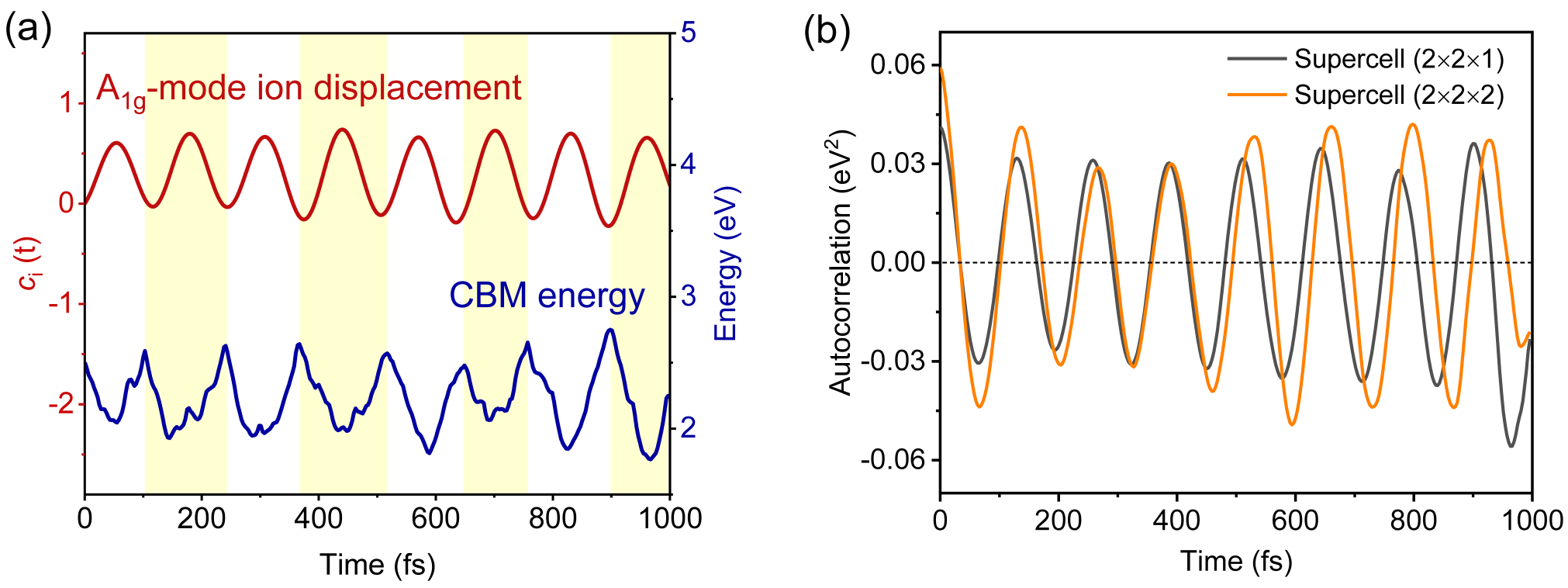

**Fig. 3. Temporal correlation between the Γ-point $A_{1g}$ breathing motion and CBM dynamics.** (a) Temporal evolution of the projected Γ-point $A_{1g}$-mode coefficient $c_i(t)$ and the CBM energy obtained from a 300 K AIMD simulation using a supercell 2×2×2 containing 904 atoms. The two quantities exhibit a clear cycle-by-cycle correspondence. (b) Autocorrelation functions of the CBM energy obtained from the 2×2×1 and 2×2×2 AIMD simulations, both showing a pronounced oscillatory component with a characteristic period of approximately 127 fs.

To directly examine the temporal relation between the $A_{1g}$ breathing motion and the CBM dynamics, we projected the AIMD ionic displacements onto the normalized phonon eigenvectors according to $\mathbf{Q}(\mathrm{t}) \approx \sum_{\mathrm{i}=1}^{\mathrm{n}} \mathrm{c}_{\mathrm{i}}(\mathrm{t}) \cdot \mathbf{q}_{\mathrm{i}}$, where $\mathbf{Q}(\mathrm{t})$ is the AIMD displacement vector, $\mathbf{q}_{\mathrm{i}}$ is the normalized eigenvector of the *i*th phonon mode, and $\mathrm{c}_{\mathrm{i}}$ is the corresponding projection coefficient obtained by least-squares fitting. As shown in Fig. 3(a), the component exhibits a clear cycle-by-cycle correspondence with the CBM energy: each breathing cycle is accompanied by a corresponding CBM modulation on essentially the same timescale. Importantly, this behavior is observed in the enlarged 2×2×2 supercell containing 904 atoms, indicating that this correspondence is not specific to the original 2×2×1 supercell. Consistently, the CBM autocorrelation functions from the two supercells exhibit closely comparable oscillatory behavior, both with a characteristic period of approximately 127 fs [Fig. 3(b)]. This shows that the characteristic periodic component is retained in the time-correlation statistics of the equilibrium CBM fluctuations rather than arising solely from visual inspection of a limited trajectory segment. Together, these results establish a close temporal correlation between the breathing coordinate and the characteristic CBM dynamics.

The close correspondence between the $A_{1g}$ breathing motion and the CBM dynamics raises an important question: why does this particular mode dominate the electronic response when multiple phonon modes are simultaneously thermally excited? To address this question, we analyzed the vibrational density of states (VDOS) and decomposed the AIMD ionic displacements into individual phonon-mode components. The VDOS was acquired by evaluating the Fourier transform of the velocity autocorrelation function based on the AIMD trajectories at 300 K [36,37]. The VDOS is given by:

$$f(\omega) = \hat{\mathcal{F}}[\gamma(t)] = \frac{1}{k_B T}\int_{-\infty}^{\infty} \frac{\langle \sum v_i(t)\cdot v_i(0)\rangle}{\langle \sum v_i^2(0)\rangle} e^{-i\omega t} dt, \tag{1}$$

where $\omega$ is the vibration frequency, $\hat{\mathcal{F}}$ is the Fourier transform operator, $k_B$ is the Boltzmann constant, $T$ is the absolute temperature, and $v_i(t)$ is the velocity of the *i*th ion at time $t$. As shown in Fig. 4(a), the octahedral motion involves four dominant phonon modes: $T_{2u}$, $T_{2g}$, $E_g$, and $A_{1g}$. Notably, the $T_{2g}$, $E_g$, and $A_{1g}$ modes coincide with the Raman-active modes identified experimentally [Fig. 2(a)], confirming that the *ab initio* simulations capture the physically relevant lattice thermal vibrations. Using the mode-projection procedure described above, we decomposed the original 2×2×1 AIMD trajectory into these four dominant components [Fig. 4(b)]. Although the $T_{2g}$ component exhibits the largest ionic displacement, the $A_{1g}$ component shows the closest temporal correspondence with the CBM energy. Thus, the selective electronic response cannot be explained simply by the relative amplitudes of the thermally excited phonons, motivating a direct evaluation of their electron–phonon coupling strengths.

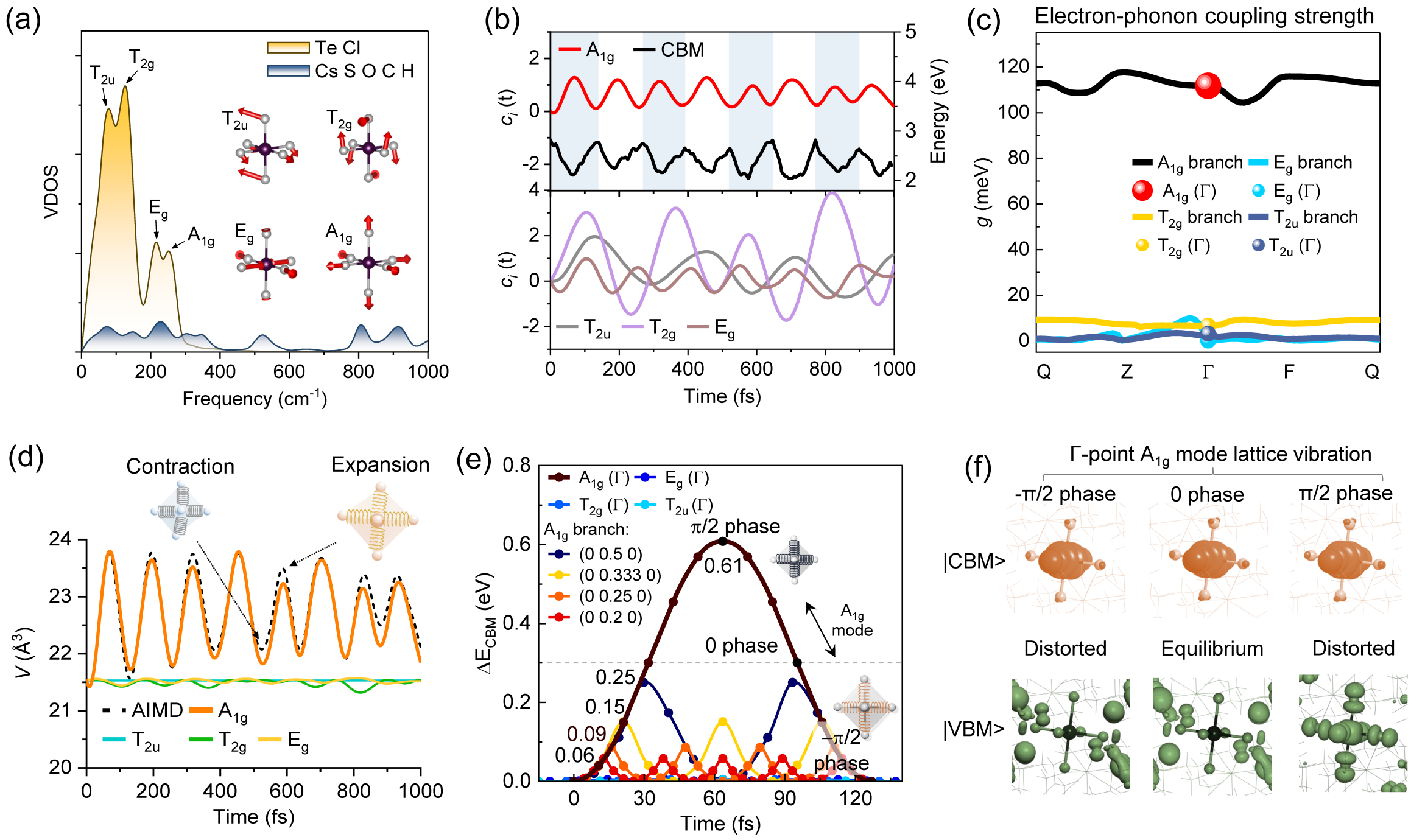

**FIG. 4. Microscopic mechanism underlying the Γ-point $A_{1g}$-driven bandgap dynamics.** (a) Atom-projected vibrational density of states (VDOS) of the DMSO crystal. (b) Temporal evolution of atomic displacements projected onto four dominant Γ-point phonon modes and the CBM energy level, obtained from the 2×2×1 AIMD trajectory. (c) First-principles electron–phonon coupling strength between the CBM state and each phonon mode along four phonon dispersion branches. (d) Temporal evolution of the octahedral volume alongside its decomposition into individual Γ-point phonon mode contributions. (e) Evolution of the CBM energy driven by each individual phonon mode over one vibrational period of Γ-point $A_{1g}$. (f) Isosurfaces of the CBM and VBM states at three representative phases ($-\pi/2$, 0, and $+\pi/2$) of the Γ-point $A_{1g}$-driven oscillation shown in (e).

We therefore evaluated the electron–phonon coupling matrix elements of these modes using the EPW (electron–phonon coupling using Wannier functions) formalism. The electron–phonon coupling matrix element $g_{mn,v}(\mathbf{k},\mathbf{q})$ represents the probability amplitude for an electron to scatter from the state $|\psi_{\mathbf{nk}}\rangle$ to $|\psi_{\mathbf{mk+q}}\rangle$ by absorbing a phonon of $|\varphi_{\mathbf{qv}}\rangle$. It follows the following relation [38,39]:

$$g_{mn,v}(\mathbf{k},\mathbf{q}) = \left(\frac{\hbar}{2M\omega_{\mathbf{qv}}}\right)^{1/2} \langle\psi_{\mathbf{mk+q}}|\, \partial_{\mathbf{qv}}V|\psi_{\mathbf{nk}}\rangle, \tag{2}$$

where M and $\omega_{\mathbf{qv}}$ are the atomic mass and the angular frequency of the phonon $|\varphi_{\mathbf{qv}}\rangle$, $\partial_{\mathbf{qv}}V$ is the derivative of the self-consistent potential energy associated with a phonon of wavevector $\mathbf{q}$, branch index v, and frequency $\omega_{\mathbf{qv}}$. As shown in Fig. 4(c), the $A_{1g}$-like branch exhibits much stronger electron–phonon coupling than the other three branches throughout the sampled wavevector path, with $g \approx 115$ meV for the Γ. In comparison, the Γ-point $E_g$, $T_{2g}$, and $T_{2u}$ modes have coupling strengths below 6.6 meV. This pronounced branch selectivity can be understood from the spatial correspondence between the localized CBM and the lattice distortions. The CBM is predominantly localized on the inorganic $[TeCl_6]^{2-}$ octahedra, while the $A_{1g}$ mode produces a fully symmetric expansion and contraction of the octahedral cage, strongly perturbing the local electronic environment of the CBM. By contrast, the $E_g$, $T_{2g}$, and $T_{2u}$ modes involve nonsymmetric distortions and yield much smaller coupling matrix elements. Consistent with this picture, decomposition of the octahedral-volume dynamics along the AIMD trajectory shows that the $A_{1g}$ component makes the dominant contribution to the temporal variation of the $[TeCl_6]^{2-}$ octahedral volume [Fig. 4(d)]. These results identify the $A_{1g}$-like branch as the dominant microscopic electron–phonon coupling channel for the flat conduction-band edge.

A crucial insight emerges when comparing microscopic electron–phonon coupling with the resulting CBM modulation. As shown in Fig. 4(c), finite-q modes on the $A_{1g}$-like branch can still exhibit appreciable coupling to the flat-band CBM, yet the induced CBM shifts remain substantially smaller than that of the Γ-point $A_{1g}$ mode [Fig. 4(e) and Fig. S6]. Thus, a large microscopic coupling strength alone is not sufficient to generate a comparably large macroscopic band-edge response. The additional requirement is real-space phase selectivity. The Γ-point $A_{1g}$ mode is a unit-cell-synchronous breathing distortion in which all octahedra expand and contract in phase. By contrast, finite-wavelength modes on the same branch contain intercell phase shifts that generate nonuniform spatiotemporal phase patterns (Fig. S7). These phase patterns lead to a systematic wavelength dependence of the CBM response. For modes with wavelengths of 2a, 3a, 4a, and 5a, represented by q = (0, 0.5, 0), (0, 0.333, 0), (0, 0.25, 0), and (0, 0.2, 0), respectively, the apparent CBM recurrence period evolves from T/2 to T/3, T/4, and T/5, while the modulation amplitude decreases from about 0.25 to 0.06 eV [Fig. 4(e) and Fig. S6], where a denotes the lattice constant along the corresponding propagation direction. Thus, longer-wavelength finite-q components produce progressively smaller and more rapidly recurring CBM modulations despite retaining appreciable microscopic coupling.

The finite-q components most likely to perturb the Γ-point pulsation are therefore the two-cell-periodic Brillouin-zone-boundary modes, which exhibit the largest finite-q CBM modulations and correspond to the shortest real-space periodicities along the relevant lattice directions. In particular, the F (0, 0.5, 0), Z (0, 0, 0.5), and Q (0, 0.5, 0.5) modes yield modulations of approximately 0.25–0.27 eV. All such boundary wavevectors are explicitly sampled by the 2×2×2 AIMD supercell. Their CBM responses, however, recur every T/2 and therefore do not provide a counter-phase fundamental-T component capable of cancelling the Γ-point response. Consistently, the phase-sampled superposition analysis in Fig. S8 shows that varying the relative phases of the dominant boundary components changes the detailed waveform but leaves the fundamental-period pulsation associated with the Γ-point $A_{1g}$ component visible throughout the sampled phase space. Together with the weak coupling of the other phonon branches, these results establish a dual selection mechanism for the giant bandgap pulsation: reciprocal-space branch selectivity, which makes the $A_{1g}$-like breathing branch the dominant microscopic coupling channel, and real-space phase selectivity, which makes the unit-cell-synchronous Γ-point component the dominant source of the macroscopic CBM pulsation. As illustrated in Fig. 4(f), the Γ-point $A_{1g}$ breathing distortion strongly modulates the CBM state and is accompanied by a marked change in the VBM character. Accordingly, neither modes on other phonon branches nor finite-q modes on the $A_{1g}$-like branch can generate CBM modulations comparable to that of the Γ-point $A_{1g}$ mode.

The remarkably large bandgap variation observed in the AIMD trajectory (~0.95 eV peak-to-peak) warrants quantitative justification. The average CBM modulation amplitude driven by the Γ-point $A_{1g}$ breathing mode is approximately 0.61 eV. According to our single-mode structural driving calculations, a $TeCl_6$-octahedron-averaged mass-weighted root-mean-square (RMS) displacement amplitude of only 0.036 Å is sufficient to reproduce this value [Fig. 4e]. This corresponds to individual Cl displacement amplitudes of approximately 0.037–0.049 Å, consistent with the scale of thermal ionic motion observed in the 300 K AIMD trajectory (Fig. S5), and therefore does not require an anomalously large structural distortion. Moreover, this giant electronic response is highly mode-selective: when the same 0.036 Å amplitude is applied to other phonon branches, the resulting CBM modulations are at least one order of magnitude smaller [Fig. 4e]. Thus, the large band-edge response reflects the strong sensitivity of the localized flat-band CBM to the unit-cell-synchronous $A_{1g}$ breathing coordinate rather than a large ionic displacement.

Regarding experimental observability, the predicted modulation corresponds to a collective synchronized shift of the entire flat-band edge, rather than a stochastic broadening of individual states. The $A_{1g}$ mode periodically modulates the average crystal-field potential of the inorganic octahedral sublattice. Consequently, such a mode-locked collective shift of the band edge can remain detectable as a dynamic spectral drift in time-resolved optical measurements, even in the presence of thermal broadening and excitonic linewidths, because the modulation is strongly correlated with a single structural coordinate.

The large-amplitude periodic CBM dynamics have significant implications for the optical response. We computed the time-resolved photon absorption spectrum from 200 to 600 nm accompanying the lattice evolution [Fig. 5(a)]. Both the absorption onset and the dominant absorption peak exhibit pronounced periodic shifts that closely follow the CBM evolution. In addition to these energy shifts, a distinct dip region present at the t_max configuration nearly disappears at t_min [Fig. 5(b)]. To identify the microscopic origin of these changes, we evaluated the squared transition dipole matrix elements for all relevant valence-to-conduction transitions within the energy windows corresponding to the absorption peak and valley. The occupied and unoccupied states shown in Fig. 5(b) [top and right panels] represent the dominant optical transitions, determined by the largest matrix-element contributions. The quantum transition rate (proportional to squared dipole-moment-transition matrix element and excitation light intensity) between two electronic states in the presence of $n_\lambda$ photons obeys [40]

$$W_{vc} = \frac{2\pi}{\hbar}\sum_\lambda \left(\frac{\hbar}{2\varepsilon_0 n^2 V\omega}\right)\left(\frac{e}{m_0}\right)^2 |\langle\varphi_c|\exp(i\mathbf{k}_\lambda\cdot\mathbf{r})\boldsymbol{\epsilon}_\lambda\cdot\mathbf{p}|\varphi_v\rangle|^2 n_\lambda \delta(E_c - E_v - \hbar\omega), \quad (3)$$

where $|\varphi_v\rangle$ (eigenenergy: $E_v$) and $|\varphi_c\rangle$ (eigenenergy: $E_c$) are initial valence band (VB) state and final conduction band (CB) state, $\omega$ is the photon angular frequency, and the summation is over photon modes with wavevector $\mathbf{k}_\lambda$ and unit polarization vector $\boldsymbol{\epsilon}_\lambda$. $m_0$, $\mathbf{r}$, $\mathbf{p}$ are electron mass, position vector, and canonical momentum. V and n are crystal volume and refractive index. $\varepsilon_0$ is the vacuum dielectric constant. The electron jumps from the VB state of wavevector $\mathbf{k}_v$ to the CB state of wavevector $\mathbf{k}_c$ via absorption of a photon of wavevector $\mathbf{k}_\lambda$. The analysis reveals that the absorption dip originates from inter-sublattice transitions, whereas the strongest peak corresponds to intra-sublattice transitions. Consistent with this picture, the electronic band structures and PDOS at t_max and t_min reveal a pronounced temporal redistribution of orbital character [Figs. 5(c)–(5(e)]. The VBM state undergoes a clear switching from organic-dominated to inorganic-dominated character between these configurations, consistent with the charge-density evolution shown in Fig. 4(f). These results indicate that while the CBM dynamics controls the evolution of absorption energies, the concomitant modulation of the valence-band orbital character governs the absorption intensity, yielding a fully dynamic and mode-selective optical response.

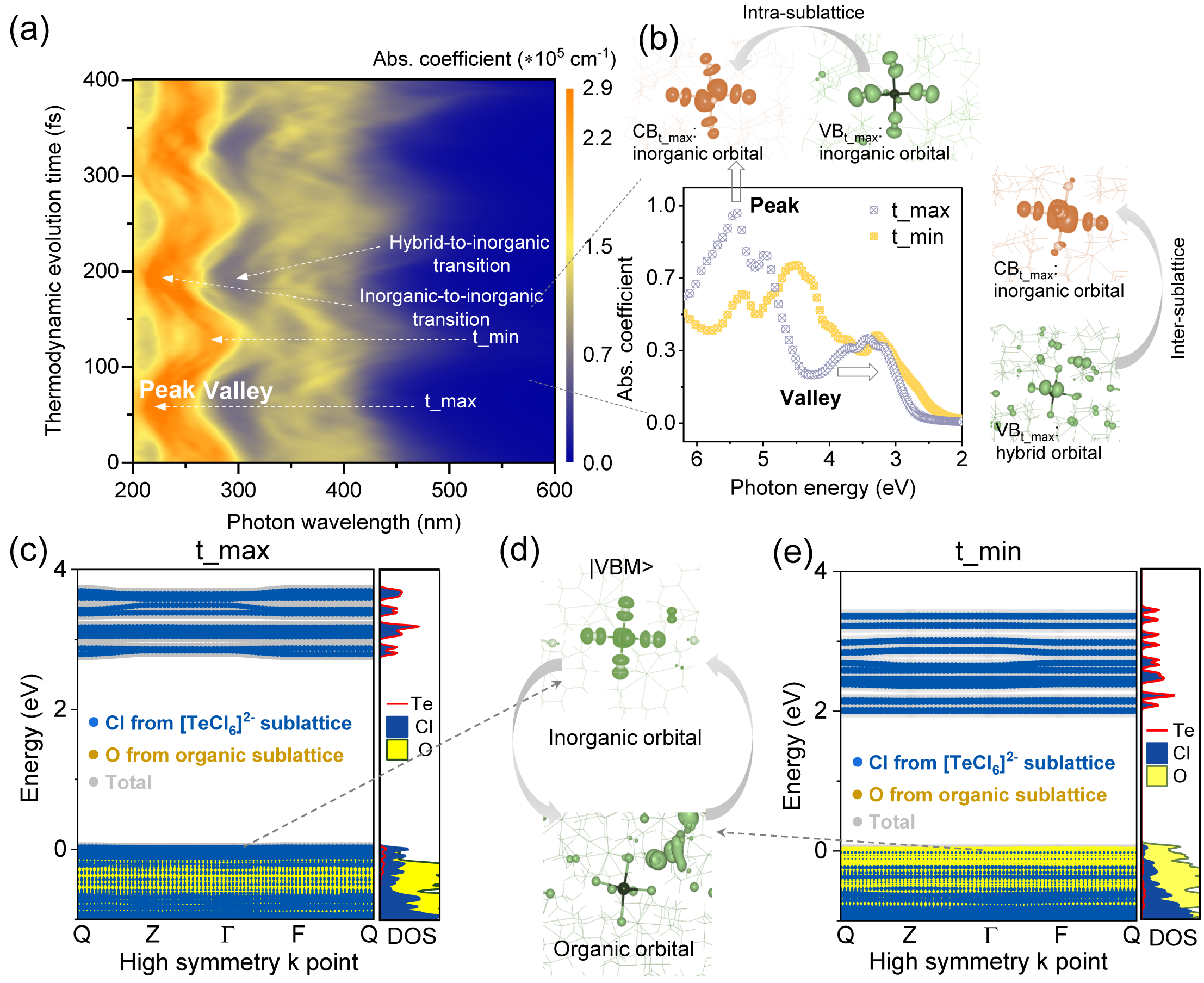

**Fig. 5. Time-dependent photon absorption and dynamic orbital hybridization in the DMSO crystal.** (a) Two-dimensional contour plot showing the temporal evolution of the photon absorption spectrum of the DMSO crystal along a segment of the AIMD trajectory at 300 K. The absorption spectra were computed within the independent-particle approximation using the frequency-dependent dielectric function, and explicitly include the optical transition matrix elements implemented in the PAW formalism. (b) Absorption coefficient as a function of photon energy at the marked t_max and t_min time points in (a). Top and right panels show isosurfaces of the conduction band and valence band states that contribute most to the photon absorption at the corresponding peak and valley points. (c, e) Electronic structure and projected

density of states at t_max and t_min, respectively. (d) Isosurfaces of the VBM states at these two time points. Panels (c–e) explicitly reveal a dynamic orbital hybridization between the inorganic $[TeCl_6]^{2-}$ and organic sublattices.

It is important to discuss the experimental observability of the predicted bandgap dynamics. In a real crystal, the phase of the Γ-point $A_{1g}$ breathing mode may vary across independent crystalline domains. If an optical measurement averages over many such domains with uncorrelated local phases, the macroscopic signal would indeed be partially smeared out, which could weaken the phase-resolved oscillatory component. Therefore, the predicted effect is best addressed using spatially resolved techniques that minimize ensemble averaging. Recent advances in ultrafast nano-optics and scanning probe spectroscopy have made it possible to resolve local electronic dynamics with sub-micron and even atomic spatial resolution [41]. Specifically, pump–probe micro-spectroscopy, time-resolved cathodoluminescence, and lightwave scanning tunneling microscopy are emerging as powerful tools to probe phonon-driven electronic modulations within individual domains. Under such domain-selective or spatially resolved measurement conditions, the pronounced Γ-point $A_{1g}$-induced bandgap pulsation should remain observable despite the presence of local disorder or domain boundaries. Moreover, since the characteristic period of the bandgap pulsation (~127 fs) is much shorter than typical phonon lifetimes (on the picosecond scale), the effect is expected to manifest over multiple cycles before anharmonic phonon–phonon scattering significantly diminishes the correlated response.

We note that the present calculations of the optical absorption are performed within the independent-particle approximation and therefore do not explicitly include excitonic effects. A quantitative treatment of excitonic effects would require, for example, GW–BSE calculations for multiple thermally distorted structures, which is computationally prohibitive for the present large hybrid system. However, because the primary finding of this work is the periodic modulation of the single-particle band-edge energies, the predicted modulation frequency and the qualitative trend of the optical shift are expected to remain robust. Excitonic effects may modify the absolute absorption strength and lineshape, but they are not expected to alter the fundamental bandgap dynamics driven by the Γ-point $A_{1g}$ breathing mode.

In summary, we have uncovered a counterintuitive regime in which the electronic bandgap exhibits giant terahertz dynamics under thermal equilibrium. This behavior arises from a dual selection mechanism: reciprocal-space branch selectivity makes the $A_{1g}$-like breathing branch the dominant microscopic electron–phonon coupling channel, while real-space phase selectivity favors the unit-cell-synchronous Γ-point hotspot component as the dominant source of the fundamental-period CBM pulsation. Finite-q modes on the same branch retain appreciable microscopic coupling but generate smaller shorter-recurrence-period responses because of their intercell phase patterns, whereas other phonon branches couple much more weakly to the flat-band edge. These results reveal how an unexpectedly ordered electronic response can emerge from intrinsically disordered thermal lattice fluctuations and suggest dual branch-and-phase selectivity as a possible route toward intrinsic ultrafast functionality in flat-band materials.

*Acknowledgments*—This work was supported by the National Natural Science Foundation of China No. 12274076.

*Data availability*—The data that support the findings of this article are openly available.

Supplemental Material

# Giant Bandgap Pulsation Driven by Hotspot Breathing Phonons in a Flat-Band Solid

Wenjie Liu,[1] Huaxin Wu,[1] and Jiyang Fan[1,2*]

[1]School of Physics, Southeast University, Nanjing 211189, P. R. China
[2]Key Laboratory of Quantum Materials and Devices of Ministry of Education, Southeast University, Nanjing 211189, P. R. China
*Contact author: jyfan@seu.edu.cn

**Calculation method**

The structural and electronic properties were calculated using the projector augmented wave (PAW) (1) method as implemented in the Vienna *ab initio* simulation package (VASP) (2–4). The exchange–correlation functional was treated within the Perdew–Burke–Ernzerhof (PBE) approximation (5). For static electronic structure, the hybrid Heyd–Scuseria–Ernzerhof (HSE06) functional (6) was further employed to confirm the flat-band character of the conduction band minimum (CBM) (Fig. S7). A plane-wave energy cutoff of 500 eV was employed. Structural optimization was conducted using a Γ-centered 4×4×3 k-point mesh for Brillouin zone sampling. Phonon density of states was computed via the PHONOPY code (7). An *ab initio* molecular dynamics (AIMD) simulation was performed in the canonical (NVT) ensemble using a 2×2×1 supercell containing 452 atoms. This supercell size was verified to be converged, as the alternative larger 2×2×2 supercell containing 904 atoms yields consistent CBM oscillation behavior [Fig. 3(a)]. A Nosé–Hoover thermostat was employed with a time step of 1 fs (8–10). Raman spectra were calculated using the Phonopy-Spectroscopy package (11). Phonon properties were further evaluated within density functional perturbation theory (DFPT) (12), as implemented in the Quantum ESPRESSO package (13). The electron–phonon coupling strength was studied by performing the Wannier-function interpolation within the EPW package (14,15). The accuracy of the Wannier interpolation was validated by comparing the DFT and Wannier-interpolated band structures. For these calculations, coarse q and k meshes of 3×3×3 and 3×3×3 were adopted for phonon and electron eigenvalues, respectively. The autocorrelation function of the CBM energy was computed to quantitatively analyze the temporal correlations in the bandgap dynamics and identify characteristic timescales. The CBM energy trajectory $E_{CBM}(t)$ was extracted from the AIMD simulation performed at room temperature. The trajectory has a total length of 1000 steps with a time step of 1 fs, covering a total simulation time of 1 ps. By subtracting the time-averaged CBM energy over the entire trajectory, we get $\delta E(t) = E_{CBM}(t) - \langle E_{CBM} \rangle$. The autocorrelation function, normalized by the variance, is defined as: $R(\tau) = \langle \delta E(t)\cdot\delta E(t+\tau)\rangle_t / \langle \delta E(t)^2 \rangle_t$. The time-resolved photon absorption spectra shown in Fig. 5(a) were calculated using the optical-transition formalism implemented in VASP. Within the independent-particle approximation, the frequency-dependent dielectric function and the resulting absorption coefficient were evaluated from interband transitions, explicitly including optical transition matrix elements within the PAW formalism. Therefore, the matrix-element contribution to the absorption intensity is directly reflected in the calculated spectra. We note that the finite supercell size in AIMD simulations imposes a cutoff on the maximum wavelength of phonons due to periodic boundary conditions.

However, the giant bandgap dynamics reported here are driven exclusively by the Γ-point (q = 0) component of the $A_{1g}$ optical branch. Because this mode corresponds to an effectively infinite wavelength ( $\lambda \to \infty$) with unit-cell-synchronous in-phase breathing, the finite supercell cutoff does not affect the physical character of this specific driver. Moreover, our mode-resolved calculations show that the finite-q components most likely perturbing the Γ-point pulsation are the two-cell-periodic Brillouin-zone-boundary modes, which exhibit the largest finite-q CBM modulations and correspond to the shortest real-space periodicities along the relevant lattice directions. Importantly, all such modes are explicitly sampled by the 2×2×2-supercell AIMD simulation.

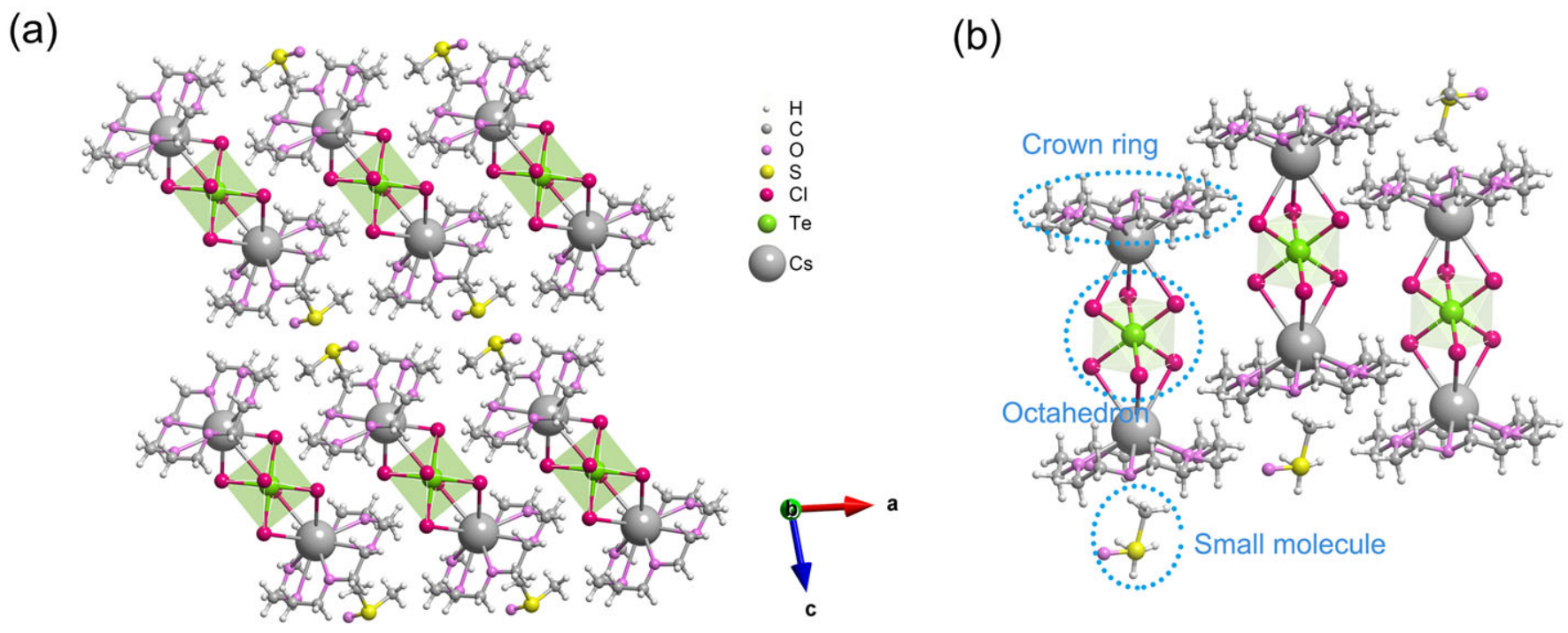

**FIG. S1. Crystal structure of the hybrid DMSO crystal.** (a) The structure features a hierarchically self-assembled architecture composed of dumbbell-like units. (b) Compositional inorganic (octahedron) and organic (crown ring and small molecule) units of the DMSO crystal.

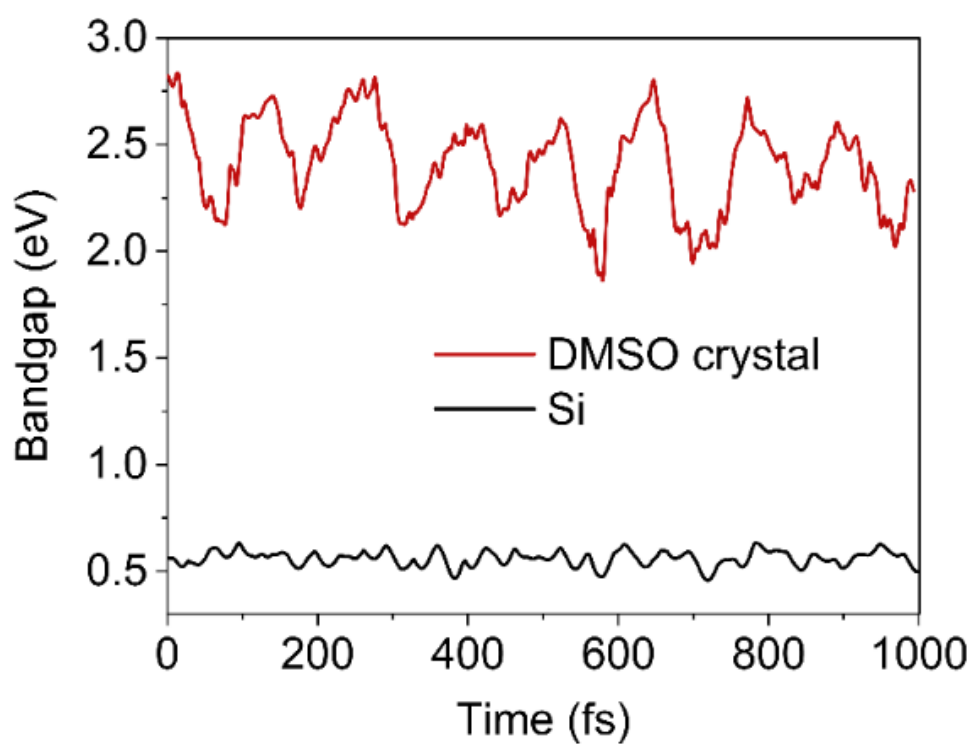

**FIG. S2. Temporal evolution of bandgaps for DMSO and Si crystals at thermal equilibrium (300 K)**. While Si exhibits typical stochastic thermal fluctuations, the DMSO crystal shows pronounced large-amplitude periodic oscillations, ruling out the possibility of numerical artifacts.

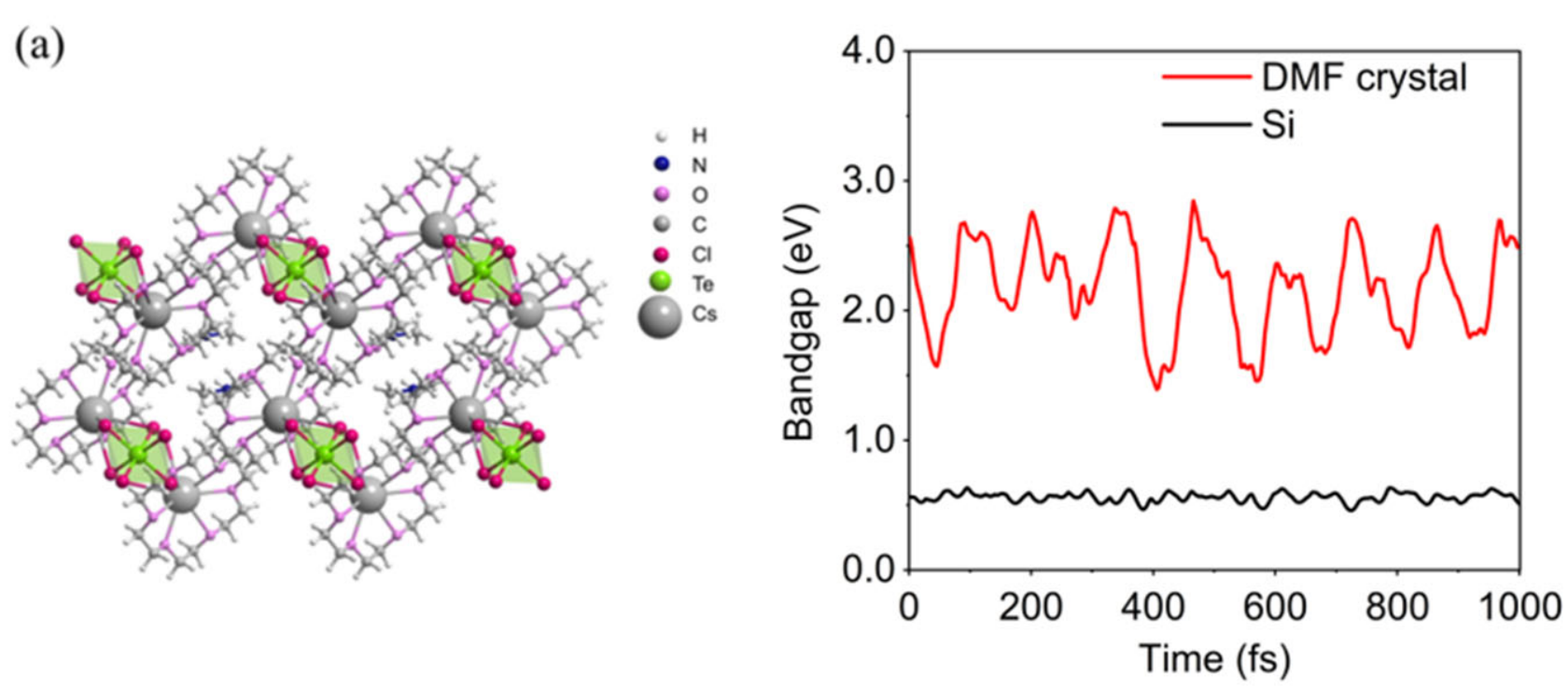

**FIG. S3. Crystal structure and similar CBM pulsations in a N,N-dimethylformamide (DMF) (substituting for DMSO)-based hybrid material.** (a) Crystal structure. (b) Time evolution of the bandgap obtained from 300 K AIMD simulations for the DMF hybrid crystal, with crystalline Si shown as a conventional semiconductor reference.

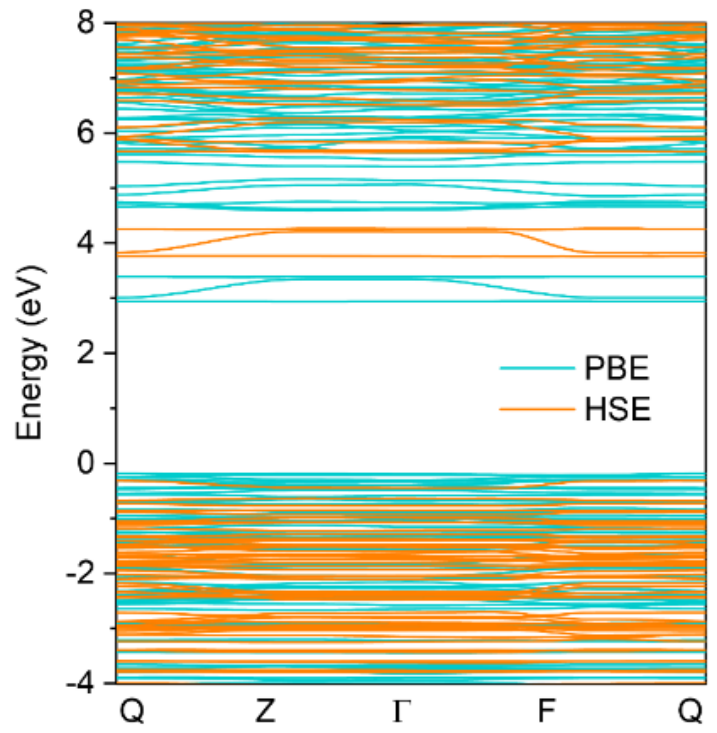

**FIG. S4. Comparison of electronic structures of the DMSO crystal adopting HSE06 and PBE exchange–correlation functionals.** Calculations using the more demanding hybrid exchange–correlation functional preserve the flat-band character.

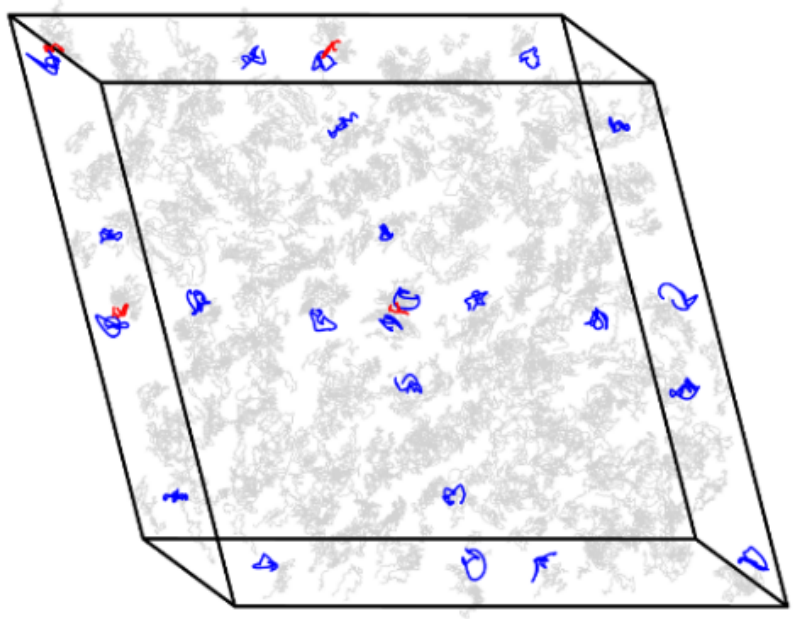

**FIG. S5. Atomic trajectories of the DMSO crystal during 1 ps AIMD simulation at 300 K.** The system was maintained at thermal equilibrium, the red, blue, and gray lines represent the trajectories of Te, Cl, and other atoms, respectively.

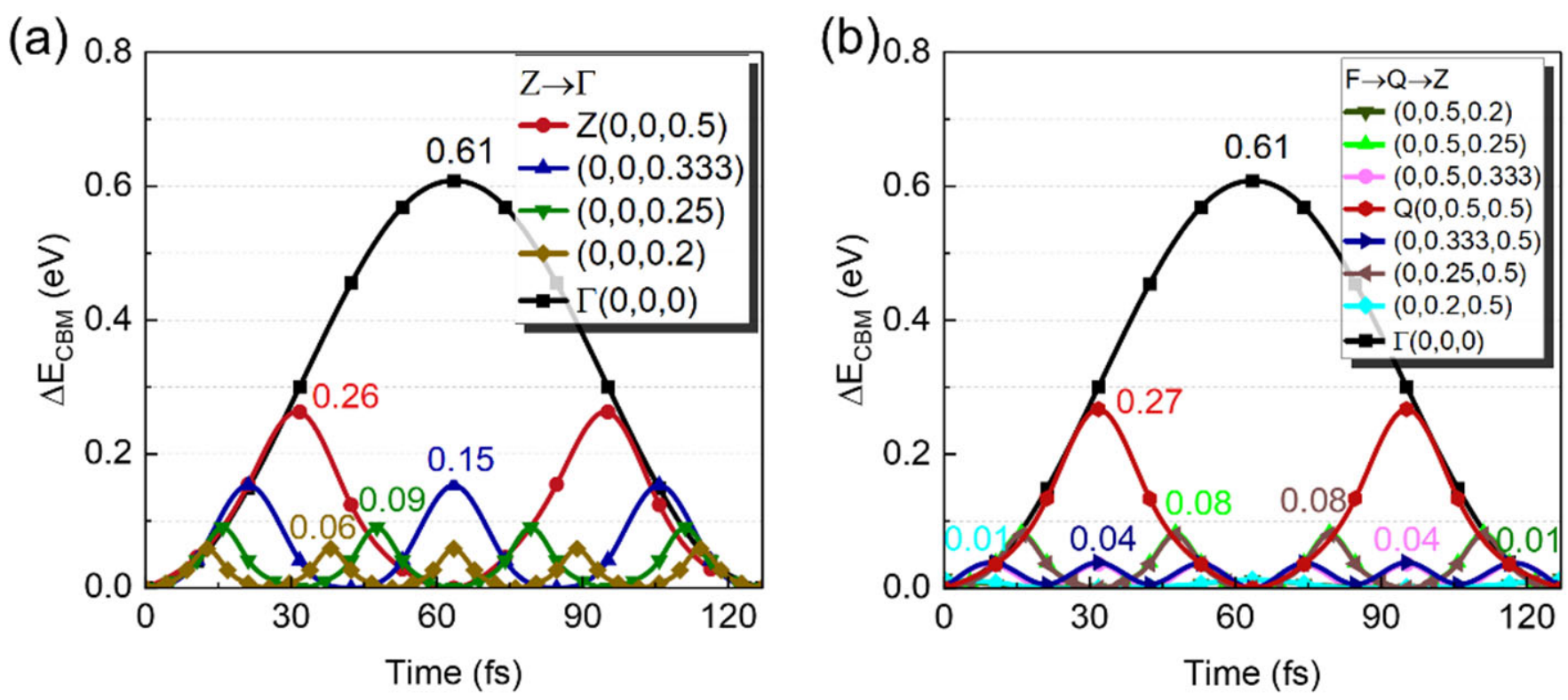

**FIG. S6. Mode-dependent CBM modulation induced by $A_{1g}$-like phonons.** (a)–(b) Evolution of the CBM energy ($\Delta E_{CBM}$) over one vibrational period obtained by imposing successive static distortions corresponding to each individual phonon mode over one vibrational cycle. For a consistent comparison, all modes were scaled to the same mass-weighted root-mean-square (RMS) displacement amplitude of 0.036 Å, evaluated over the Te and six Cl atoms of each $[TeCl_6]^{2-}$ octahedron. This amplitude was selected so that the Γ-point $A_{1g}$ mode quantitatively reproduces the average CBM modulation observed in AIMD (~0.61 eV). For finite-q modes propagating along a principal lattice direction with wavelength λ = na, the intercell phase pattern gives rise to an apparent CBM recurrence period of approximately T/n for this nearly flat phonon branch, together with a systematic reduction in modulation amplitude as the wavelength increases. For diagonal finite-q modes containing multiple nonzero wavevector components, the combined spatial phase periodicity can produce even shorter recurrence periods; for example, q = (0,0.5,0.333) and (0,0.333,0.5) exhibit an apparent recurrence of approximately T/6.

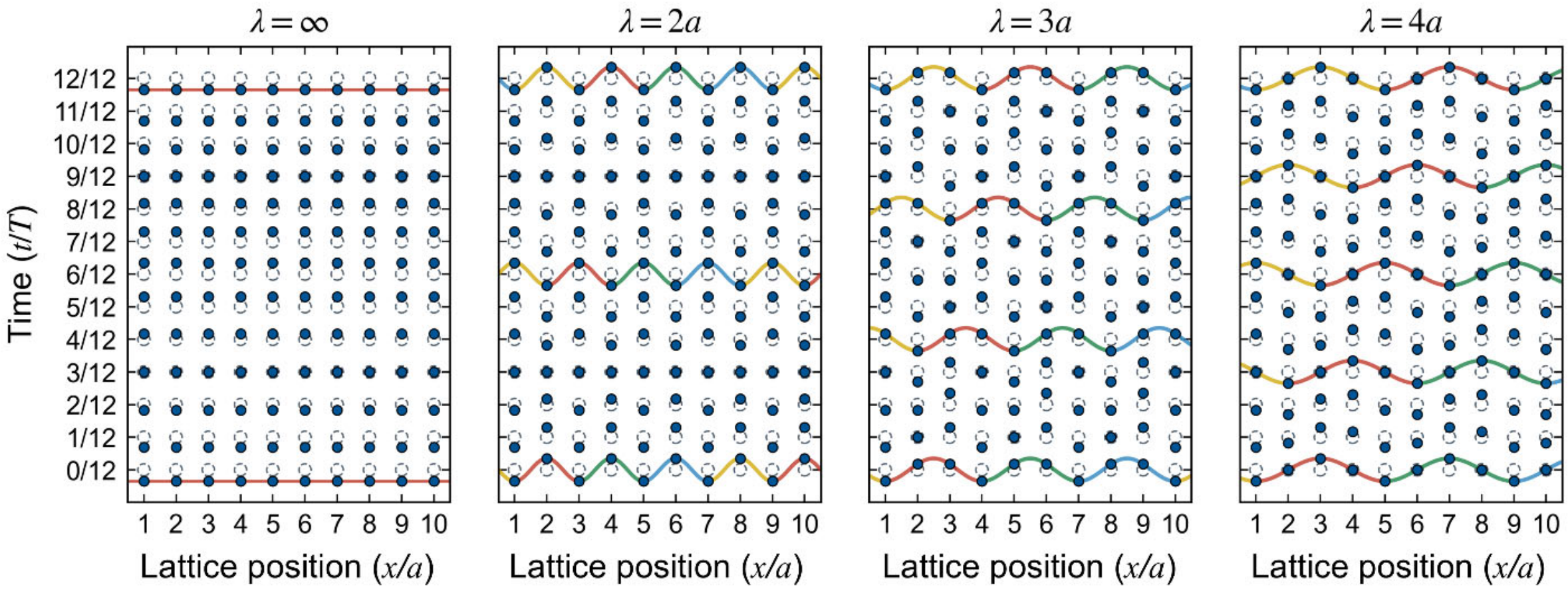

**FIG. S7. Schematic illustration of real-space phase patterns and recurrence of finite-wavelength phonons.** Lattice displacements for the Γ-point (q = 0) and representative finite-wavelength (q ≠ 0) phonon modes propagating along a principal direction. The snapshots are shown over one vibrational period T for wavelengths λ = ∞, 2a, 3a, and 4a, where a is the lattice constant. Open and solid circles denote equilibrium and instantaneous atomic positions, respectively. The Γ-point mode (λ = ∞) exhibits a spatially uniform unit-cell-synchronous displacement pattern and recovers after one full period T. By contrast, a finite-wavelength mode with λ = n×a contains an intercell phase shift of 2π/n. As the phonon evolves in time, its real-space phase pattern becomes translationally equivalent after T/n, giving rise to an apparent recurrence period of T/n in the corresponding collective CBM response; thus, the 2a, 3a, and 4a modes exhibit T/2, T/3, and T/4 recurrences, respectively. These intercell phase patterns also reduce the collective band-edge modulation relative to the unit-cell-synchronous Γ-point distortion, explaining why finite-q modes produce smaller more rapidly recurring CBM responses despite retaining appreciable microscopic electron–phonon coupling.

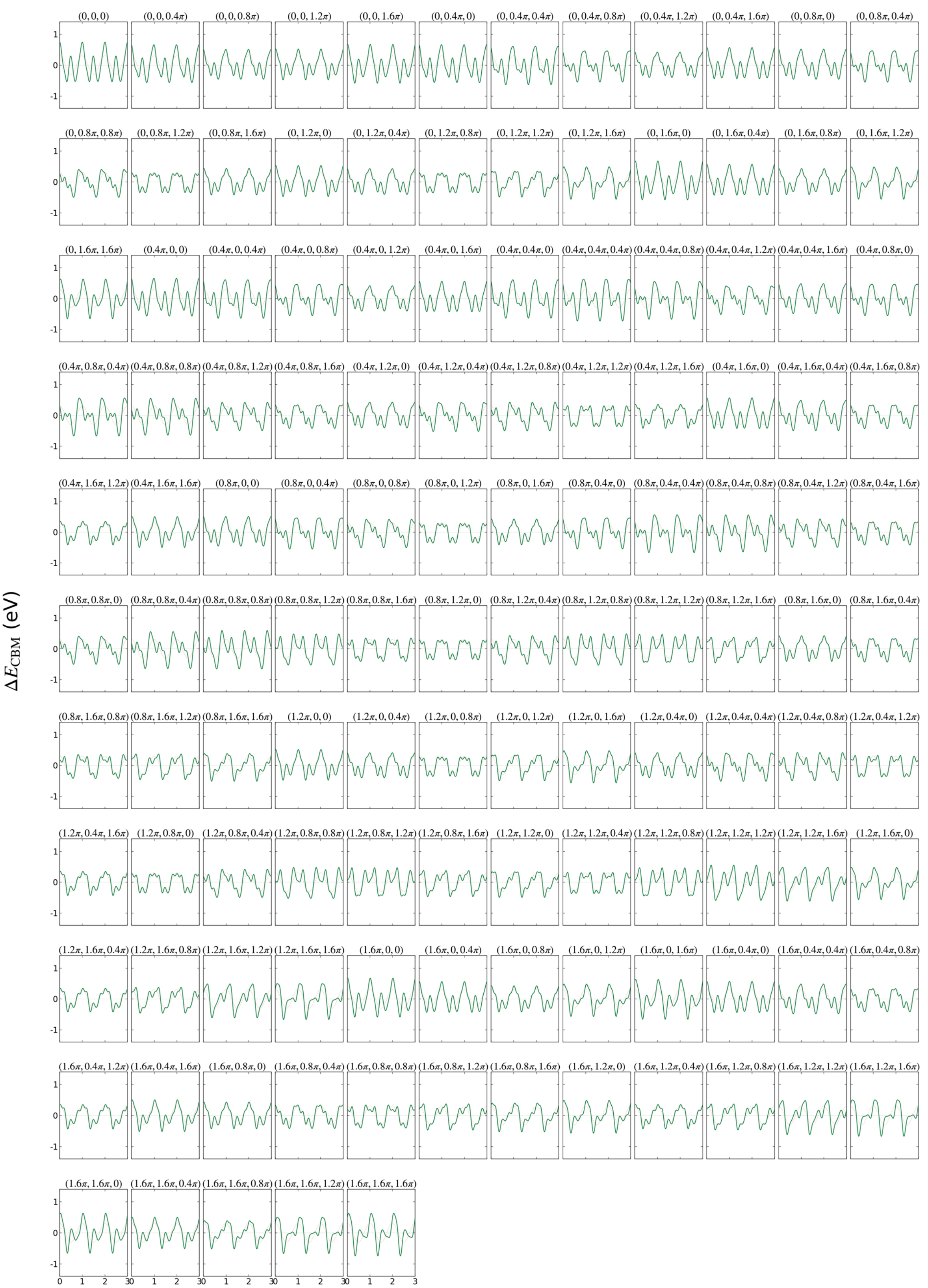
(0, 0, 0) (0, 0, 0.4π) (0, 0, 0.8π) (0, 0, 1.2π) (0, 0, 1.6π) (0, 0.4π, 0) (0, 0.4π, 0.4π) (0, 0.4π, 0.8π) (0, 0.4π, 1.2π) (0, 0.4π, 1.6π) (0, 0.8π, 0) (0, 0.8π, 0.4π)
(0, 0.8π, 0.8π) (0, 0.8π, 1.2π) (0, 0.8π, 1.6π) (0, 1.2π, 0) (0, 1.2π, 0.4π) (0, 1.2π, 0.8π) (0, 1.2π, 1.2π) (0, 1.2π, 1.6π) (0, 1.6π, 0) (0, 1.6π, 0.4π) (0, 1.6π, 0.8π) (0, 1.6π, 1.2π)
(0, 1.6π, 1.6π) (0.4π, 0, 0) (0.4π, 0, 0.4π) (0.4π, 0, 0.8π) (0.4π, 0, 1.2π) (0.4π, 0, 1.6π) (0.4π, 0.4π, 0) (0.4π, 0.4π, 0.4π) (0.4π, 0.4π, 0.8π) (0.4π, 0.4π, 1.2π) (0.4π, 0.4π, 1.6π) (0.4π, 0.8π, 0)
(0.4π, 0.8π, 0.4π) (0.4π, 0.8π, 0.8π) (0.4π, 0.8π, 1.2π) (0.4π, 0.8π, 1.6π) (0.4π, 1.2π, 0) (0.4π, 1.2π, 0.4π) (0.4π, 1.2π, 0.8π) (0.4π, 1.2π, 1.2π) (0.4π, 1.2π, 1.6π) (0.4π, 1.6π, 0) (0.4π, 1.6π, 0.4π) (0.4π, 1.6π, 0.8π)
(0.4π, 1.6π, 1.2π) (0.4π, 1.6π, 1.6π) (0.8π, 0, 0) (0.8π, 0, 0.4π) (0.8π, 0, 0.8π) (0.8π, 0, 1.2π) (0.8π, 0, 1.6π) (0.8π, 0.4π, 0) (0.8π, 0.4π, 0.4π) (0.8π, 0.4π, 0.8π) (0.8π, 0.4π, 1.2π) (0.8π, 0.4π, 1.6π)
(0.8π, 0.8π, 0) (0.8π, 0.8π, 0.4π) (0.8π, 0.8π, 0.8π) (0.8π, 0.8π, 1.2π) (0.8π, 0.8π, 1.6π) (0.8π, 1.2π, 0) (0.8π, 1.2π, 0.4π) (0.8π, 1.2π, 0.8π) (0.8π, 1.2π, 1.2π) (0.8π, 1.2π, 1.6π) (0.8π, 1.6π, 0) (0.8π, 1.6π, 0.4π)
(0.8π, 1.6π, 0.8π) (0.8π, 1.6π, 1.2π) (0.8π, 1.6π, 1.6π) (1.2π, 0, 0) (1.2π, 0, 0.4π) (1.2π, 0, 0.8π) (1.2π, 0, 1.2π) (1.2π, 0, 1.6π) (1.2π, 0.4π, 0) (1.2π, 0.4π, 0.4π) (1.2π, 0.4π, 0.8π) (1.2π, 0.4π, 1.2π)
(1.2π, 0.4π, 1.6π) (1.2π, 0.8π, 0) (1.2π, 0.8π, 0.4π) (1.2π, 0.8π, 0.8π) (1.2π, 0.8π, 1.2π) (1.2π, 0.8π, 1.6π) (1.2π, 1.2π, 0) (1.2π, 1.2π, 0.4π) (1.2π, 1.2π, 0.8π) (1.2π, 1.2π, 1.2π) (1.2π, 1.2π, 1.6π) (1.2π, 1.6π, 0)
(1.2π, 1.6π, 0.4π) (1.2π, 1.6π, 0.8π) (1.2π, 1.6π, 1.2π) (1.2π, 1.6π, 1.6π) (1.6π, 0, 0) (1.6π, 0, 0.4π) (1.6π, 0, 0.8π) (1.6π, 0, 1.2π) (1.6π, 0, 1.6π) (1.6π, 0.4π, 0) (1.6π, 0.4π, 0.4π) (1.6π, 0.4π, 0.8π)
(1.6π, 0.4π, 1.2π) (1.6π, 0.4π, 1.6π) (1.6π, 0.8π, 0) (1.6π, 0.8π, 0.4π) (1.6π, 0.8π, 0.8π) (1.6π, 0.8π, 1.2π) (1.6π, 0.8π, 1.6π) (1.6π, 1.2π, 0) (1.6π, 1.2π, 0.4π) (1.6π, 1.2π, 0.8π) (1.6π, 1.2π, 1.2π) (1.6π, 1.2π, 1.6π)
(1.6π, 1.6π, 0) (1.6π, 1.6π, 0.4π) (1.6π, 1.6π, 0.8π) (1.6π, 1.6π, 1.2π) (1.6π, 1.6π, 1.6π)
$\Delta E_{\mathrm{CBM}}$ (eV)
Time ($t/T$)

**FIG. S8. Phase-sampled superposition analysis of finite-wavelength $A_{1g}$-branch-phonon contributions to the Γ-point $A_{1g}$-related CBM pulsation.** The grid displays the time evolution of the CBM energy ($\Delta E_{CBM}$) under systematically varied initial phase combinations of the three dominant Brillouin-zone-boundary phonon modes at the F (0, 0.5, 0), Z (0, 0, 0.5), and Q (0, 0.5, 0.5) points. Their phases ($\phi_F$, $\phi_Z$, $\phi_Q$) were independently scanned from 0 to 2π with a step of 0.4π, as labeled above each panel. Additional finite-q contributions were included with random initial phases, including other boundary modes at q = (0.5,0,0), (0.5,0.5,0), (0.5,0,0.5), and (0.5,0.5,0.5), whose mode-resolved CBM modulation amplitudes are approximately 0.11, 0.09, 0.13, and 0.10 eV, respectively, together with longer-wavelength components with λ = 3a–5a along the sampled Brillouin-zone paths. Throughout the sampled phase space, the superposed finite-q contribution remains substantially smaller than the characteristic Γ-point $A_{1g}$ modulation and does not develop a comparable fundamental-T component. This phase-superposition analysis is a controlled harmonic test and does not include the full anharmonic phonon dynamics or mode–mode interactions present in AIMD; it is therefore intended to assess the phase robustness of the fundamental- response rather than to reproduce the detailed AIMD waveform. Because their CBM responses recur at T/n rather than T, they cannot provide a counter-phase fundamental-T component that directly cancels the Γ-point response through phase superposition.